\documentclass[twocolumn,aps,floatfix,amsmath,amssymb,showpacs,nofootinbib]{revtex4}
\usepackage{hyperref}
\usepackage{graphicx}

\newcommand{\be}{\begin{equation}} \newcommand{\ee}{\end{equation}}
\newcommand{\bea}{\begin{eqnarray}} \newcommand{\eea}{\end{eqnarray}}
\newcommand{\bi}{\begin{itemize}} \newcommand{\ei}{\end{itemize}}
\newcommand{\bn}{\begin{enumerate}} \newcommand{\en}{\end{enumerate}}
 
\newcommand{\ie}{i.e., }

\newcommand{\ket}[1]{\ensuremath{|{#1\rangle}}} 
\newcommand{\bra}[1]{\ensuremath{{\langle #1}|}}
\newcommand{\braket}[2]{\ensuremath{{\langle #1}|{#2 \rangle}}}
\newcommand{\ketbra}[2]{\ensuremath{|{#1 \rangle}{\langle #2}|}}

\newcounter{romnum}

\newcounter{bracketnum}

\begin{document}

\title{Self-induced decoherence approach: Strong limitations on its
  validity in a simple spin bath model and on its general physical
  relevance}

\author{Maximilian Schlosshauer}

\email{MAXL@u.washington.edu}

\affiliation{Department of Physics, University of Washington, Seattle,
  Washington 98195, USA}


\begin{abstract}
  
  The ``self-induced decoherence'' (SID) approach suggests that (1)
  the expectation value of any observable becomes diagonal in the
  eigenstates of the total Hamiltonian for systems endowed with a
  continuous energy spectrum, and that (2) this process can be
  interpreted as decoherence. We evaluate the first claim in the
  context of a simple spin bath model. We find that even for large
  environments, corresponding to an approximately continuous energy
  spectrum, diagonalization of the expectation value of random
  observables does in general not occur. We explain this result and
  conjecture that SID is likely to fail also in other systems composed
  of discrete subsystems.  Regarding the second claim, we emphasize
  that SID does not describe a physically meaningful decoherence
  process for individual measurements, but only involves destructive
  interference that occurs collectively within an ensemble of
  presupposed ``values'' of measurements. This leads us to question
  the relevance of SID for treating observed decoherence effects.

\end{abstract}

\maketitle

\vspace{1cm}

\section{Introduction}

In a series of papers \cite{Castagnino:2005:yb,Castagnino:2004:tc,Castagnino:2003:yb,Castagnino:2002:ic,%
Castagnino:2001:pq,Castagnino:2001:pl,Castagnino:2000:oq,%
Castagnino:2000:pb,Castagnino:2000:ja,Laura:1999:on,%
Castagnino:2000:vt,Castagnino:1998:tz,Laura:1998:io,Castagnino:1997:wz},
the authors claim to present a ``new approach to decoherence''
\cite{Castagnino:2004:tc}, termed ``self-induced decoherence'' (SID).
Their main assertion is that, for systems endowed with a continuous
energy spectrum, the expectation value of an observable will become
diagonal in the eigenbasis of the Hamiltonian of the system, and that
this effect can be viewed as decoherence.

The basic idea underlying SID goes back to well-known arguments in the
context of quantum measurement and the theory of irreversible
processes
\cite{Pauli:1928:gg,Kampen:1954:om,Daneri:1962:om,%
  Hepp:1972:pa,Peres:1980:im,Hove:1959:tt,Hove:1955:qz,Hove:1957:ra}.
It rests on the observation that a superposition of a large number of
terms with random phases in the expression for the expectation value
of a typical observable, or for the matrix elements of the density
operator, leads to destructive interference.  The phase differences
are either due to a random-phase assumption \cite{Pauli:1928:gg}, or,
as in SID, are created dynamically through the time evolution factor
$e^{iEt/\hbar}$ associated with each energy eigenstate in
the superposition.  These destructive interference effects are then
responsible for the diagonalization of the expectation value in the
energy eigenbasis as described by SID.

However, this process differs strongly from the mechanism of
environment-induced decoherence (EID)
\cite{Joos:2003:jh,Zurek:1981:dd,Zurek:1982:tv,Zurek:1993:pu,%
Zurek:1998:re,Zurek:2002:ii,Schlosshauer:2003:tv,Zeh:1970:yt,%
Zeh:1973:wq}. EID understands decoherence as the practically
irreversible dislocalization of local phase relations between
environment-selected preferred basis states due to entanglement with
an environment. The approximate diagonality of the expectation value
of local observables expressed in the preferred basis is only a formal
phenomenological consequence of the relative states of the environment
becoming rapidly orthogonal during the decoherence process.  The fact
that SID does not require an explicit environment interacting with the
system motivated the term ``self-induced'' and was suggested
\cite{Castagnino:2004:tc} to circumvent the question of a proper
interpretation of the concept of ``observational ignorance of the
environment'' in EID
\cite{Espagnat:1988:cf,Joos:2003:jh,Zurek:1998:re,Schlosshauer:2003:tv}.

This paper pursues two main goals. First, after formalizing the basic
idea of SID (Sec.~\ref{sec:basic-concepts}), we shall discuss the
question to what extent SID can claim to describe a physically
relevant decoherence process (Sec.~\ref{sec:differences}). In
particular, we will argue that, contrary to the claim of its
proponents \cite{Castagnino:2004:tc}, SID does not constitute a ``new
viewpoint'' on decoherence in the usual definition of EID. Second, we
shall study whether diagonalization of the expectation value of random
observables in the energy eigenbasis is obtained in the context of an
explicit spin bath model (Sec.~\ref{sec:zurek}).  Deliberately, we
have chosen a discrete model to investigate the required degree of
``quasicontinuity'' for SID to work as claimed. To anticipate, we find
that even for bath sizes large compared to what is typically
considered in EID, no general decay of off-diagonal terms is found,
unless both the observable and the initial state of the bath are
appropriately restricted.  We explain and discuss this result in
Sec.~\ref{sec:discussion}, and present our conclusions in
Sec.~\ref{sec:conclusion}.

\section{\label{sec:basic-concepts}Self-induced decoherence}

The basic formalism of SID as developed in
Refs.~\cite{Castagnino:2004:tc,%
Castagnino:2003:yb,Castagnino:2002:ic,%
Castagnino:2001:pq,Castagnino:2001:pl,Castagnino:2000:oq,%
Castagnino:2000:pb,Castagnino:2000:ja,Laura:1999:on,%
Castagnino:2000:vt,Castagnino:1998:tz,Laura:1998:io,Castagnino:1997:wz}
considers an arbitrary observable 
\be
\widehat{O} = \int dE \int dE' \, \widetilde{O}(E,E') \ket{E} \bra{E'},
\ee
expanded in the eigenstates $\ket{E}$ of the Hamiltonian $\widehat{H}
= \int dE \, E \ket{E} \bra{E}$ with continuous spectrum. In the
general treatment, only observables with
\be \label{eq:oo}
\widetilde{O}(E,E') = O(E) \delta(E-E') + O(E,E')
\ee
are considered, where $O(E)$ and the $O(E,E')$ are assumed to be
regular functions.  The time evolution of the expectation value
$\langle \widehat{O} \rangle_{\Psi(t)}$ of $\widehat{O}$ in the pure
state $\ket{\Psi(t)}=e^{-iHt}\ket{\Psi_0}=\int dE \, e^{-iEt}
\braket{E}{\Psi_0} \ket{E}$ (setting $\hbar = 1$) is then given by
\begin{multline} \label{eq:ev} \langle \widehat{O} \rangle_{\Psi(t)} = \int dE \, O(E)
|\braket{E}{\Psi_0}|^2 \\ \,\, + \int dE \int dE' \, e^{-i\Delta E t} O(E,E')
\braket{E}{\Psi_0} \braket{\Psi_0}{E'},  
\end{multline}
where $\Delta E = E-E'$.  For large $t$, the phase factor $e^{-i\Delta
  E t}$ fluctuates rapidly with $\Delta E$, which leads to destructive
interference in the double integral if the multiplying function
$f(E,E') \equiv O(E,E') \braket{E}{\Psi_0} \braket{\Psi_0}{E'}$
varies comparably slowly.  To formalize this argument, SID employs the
Riemann-Lebesgue theorem \cite{Reed:1975:un}, which prescribes that
\be \label{eq:rl}
\lim_{t \rightarrow \infty} \int dz \, g(z) e^{izt} = 0,
\ee
if $g(z)$ is a regular function and $L^1$ integrable (\ie $\int dz \,
|g(z)| < \infty$). Provided these conditions are satisfied by
$f(E,E')$, it is concluded that
\be\label{eq:ev-d}
 \langle \widehat{O} \rangle_{\Psi(t)}
\longrightarrow \int dE \, O(E) |\braket{E}{\Psi_0}|^2,
\ee
for large $t$. Thus, the off-diagonal terms $E \not= E'$ have
collectively disappeared, which in SID is interpreted as ``decoherence
in the expectation value.'' Formally, the SID program introduces a
``diagonal-equivalent'' density matrix $\rho_d$,
\be \label{eq:rhod}
\rho_d = \int dE \, |\braket{E}{\Psi_0}|^2 \ketbra{E}{E},
\ee 
which satisfies $\langle \widehat{O} \rangle_{\rho_d} \equiv \lim_{t
  \rightarrow \infty} \langle \widehat{O} \rangle_{\rho(t)}$.  Note
that $\rho_d$ is only a formal equivalent and is not obtained through
any dynamical process. Also, expectation values of a nonexhaustive set
of observables [see Eq.~\eqref{eq:oo}] do not uniquely determine the
density matrix.  Therefore, one must not derive any conclusions about
the possibility for certain states of the system from $\rho_d$.

To summarize, the main result Eq.~\eqref{eq:ev-d} has been obtained
from two key assumptions: (1) The energy spectrum of the system is
continuous; and (2) the coefficients used in expanding the initial
state and the observable in the energy eigenbasis form regular (and
integrable) functions of the energy variable.

The first requirement of a continuous energy spectrum can be viewed as
an implicit inclusion of an internal ``environment'' with an infinite
number of degrees of freedom. However, any realistic physical system
is of finite size, and therefore the energy spacing will be discrete.
An approximate suppression of off-diagonal terms as given by
Eq.~\eqref{eq:ev-d} should therefore occur also for quasicontinuous
energy spectra, \ie for small but discrete energy spacings.

The regularity assumption (2) is crucial, since it ensures that the
phase factors $e^{i\Delta E t}$ are able to lead to the required
destructive interference of the expansion coefficients for large
times.  However, especially in the realistic case of systems of finite
size where the expansion coefficients will be a finite set of discrete
values, this condition will not hold. It is therefore important to
understand the physical meaning and the consequences of a violation of
this assumption.

Note also that the strict mathematical limit $t \rightarrow \infty$
employed in the Riemann-Lebesgue theorem, Eq.~\eqref{eq:rl}, is not
physically meaningful, and approximate suppression must therefore
occur already over finite time scales, as indicated in
Eq.~\eqref{eq:ev-d}.  Also, for the realistic case of only
quasicontinuous (i.e., essentially discrete) energy spectra, no
conclusions about an ``irreversibility'' of the decay should be
derived from the limit $t \rightarrow \infty$ (as it is done, for
example, in Ref.~\cite[p.~88]{Castagnino:2004:tc}), since the
off-diagonal terms will return to their initial values within a finite
recurrence time scale.

The issues outlined above will be illustrated and investigated in the
context of a particular model system in Sec.~\ref{sec:zurek}.

\section{\label{sec:differences}Does SID describe decoherence?}

Despite the fact that SID and EID share the term ``decoherence'' in
their name, we shall demonstrate in this section that their
foundations, scope, and physical implications are fundamentally
different.\footnote{The author is indebted to H.-D.~Zeh and E.~Joos
  for drawing strong attention to this point.}  Keeping these
differences in mind is very important for a proper interpretation of
the study of the bath model described in the following
Sec.~\ref{sec:zurek}.

As already briefly outlined in the Introduction, the standard approach
of environmental decoherence \cite{Joos:2003:jh,Zurek:1981:dd,Zurek:1982:tv,Zurek:1993:pu,%
Zurek:1998:re,Zurek:2002:ii,Schlosshauer:2003:tv,Zeh:1970:yt,%
Zeh:1973:wq} describes the consequences of the ubiquitous interaction
of any system with its environment. This leads to entanglement between
the system and the environment and singles out a preferred basis of
the system that is dynamically determined by the Hamiltonian governing
the interaction.  The relative environmental states associated with
these preferred states rapidly approach orthogonality (i.e.,
macroscopic distinguishability). Phase relations between the preferred
states that were initially associated with the system alone are now
``dislocalized'' into the system-environment combination due to the
entanglement, which constitutes the decoherence process.  In this
sense, interference between the preferred states becomes locally
suppressed, i.e., decoherence leads \emph{locally} to a transition
from a superposition to an \emph{apparent} (``improper''
\cite{Espagnat:1988:cf}) ensemble. This can be used to define
dynamically independent relative local wave-function components that
can be related to local quasiclassical properties, thereby mimicking
an \emph{apparent} ``collapse'' of the wave function
\cite{Zeh:1970:yt,Zeh:1973:wq,Zeh:1993:lt,Zeh:2000:rr,Zurek:1998:re,%
Zurek:1993:pu,Zurek:2004:yb,Zurek:2002:ii,Schlosshauer:2003:tv,Joos:2003:jh}.

The interaction between the system and its environment, often referred
to as a ``continuous measurement by the environment,'' is
observer independent and can be formulated entirely in terms of wave
functions, without reference to presumed (classical) concepts such as
``values of observables'' and expectation values (see, for example,
Chap.~2 of Ref.~\cite{Joos:2003:jh}). As it has been emphasized
frequently \cite{Espagnat:1988:cf,Schlosshauer:2003:tv,Joos:2003:jh},
the formalism of local (``reduced'') density matrices and expectation
values presupposes the probabilistic interpretation of the wave
function and ultimately relies on the occurence of a ``collapse'' of
the wave function at some stage (or on the description of an
observationally equivalent ``branching'' process in a relative-state
framework
\cite{Zeh:1970:yt,Zeh:1973:wq,Zeh:1993:lt,Zeh:2000:rr,Zurek:1998:re,%
  Zurek:1993:pu,Zurek:2004:yb,Zurek:2002:ii,Schlosshauer:2003:tv}).
The approximate diagonalization of the reduced density matrix
$\rho_\mathcal{S} = \text{Tr}_\mathcal{E} \rho_\mathcal{SE}$
(describing the probability distribution of outcomes of measurements
on the ``system $\mathcal{S}$ of interest'' immersed into an
environment $\mathcal{E}$) in the environment-selected basis should
therefore be considered only as a phenomenological consequence of EID,
but not as its essence (see also Ref.~\cite[p.~1800]{Zurek:1998:re}).
Given an ensemble of results of measurements of a local observable
$\widehat{O}_\mathcal{S}$, the suppression of off-diagonal terms in
$\rho_\mathcal{S}$ can then be related to the approximate diagonality
of the expectation value of $\widehat{O}_\mathcal{S}$ in the preferred
basis, since $\langle \widehat{O}_\mathcal{S}
\rangle_{\rho_\mathcal{SE}} = \text{Tr}_\mathcal{S} (\rho_\mathcal{S}
\widehat{O}_\mathcal{S})$.
 
In contrast with EID, SID focuses solely on the derivation of a
suppression of off-diagonal terms (in the energy eigenbasis only) in
the expectation value of observables pertaining to a single undivided
closed system; entanglement through interactions between subsystems
plays no role in SID.  As indicated earlier, the damping effect is due
to destructive interference between a large number of terms with
dynamically induced phase differences. Thus it is only the averaging
process contained in the concept of expectation values that leads to a
disappearance of interference terms. Individually, each term remains
present at all times and is not suppressed independently of the other
terms. The fact that collectively the off-diagonal terms may lead to a
mutual canceling-out must not be misinterpreted as implying that the
measurement ``outcomes'' corresponding to these terms do not occur.
Thus SID cannot pertain to the relevant problem of a loss of
interference in individual measurements. In view of this argument, the
concept of the ``diagonal-equivalent density matrix'' $\rho_d$, as
introduced by the SID program [see Eq.~\eqref{eq:rhod}], is rather
misleading, since it gives the incorrect impression of an absence of
interference terms $\ketbra{E}{E'}$, while the corresponding terms in
the expression for the expectation value are individually present at
all times.  Derivations of a ``classical limit'' based on $\rho_d$
\cite{Castagnino:2003:yb} appear to have overlooked this issue.

While SID rests on the concept of expectation values, i.e., of
weighted averages over an ensemble of measurement outcomes, it does
not explain the physical origin of these outcomes and their ensembles.
In contrast with EID, SID does not contain a dynamical account of the
measurement process itself that could motivate explanations for how
measurement outcomes arise (if only, as in EID, in an ``apparent,''
relative-state sense).  Consequently, the assumption of an \emph{a
  priori} existence of an ensemble of measurement outcomes, as it is
inherent in SID, could be viewed as a particular application of the
Copenhagen interpretation.  One might then argue that in this case
decoherence would not even be necessary in explaining the observed
absense of (macrosopic) interference effects.

Note that EID makes crucial use of the concept of locality in deriving
a loss of interference, since globally the quantum-mechanical
superposition remains unchanged, as required by the unitarity of the
time evolution of the total wave function. As frequently emphasized by
Zeh (e.g., in Refs.~\cite{Zeh:1970:yt,Zeh:2000:rr}) and others (see,
for example, Ref.~\cite{Landsman:1995:oi}), this locality can be
grounded in the (nontrivial) empirical insight that all observers and
interactions are intrinsically local. On the other hand, the
decomposition into a ``system of interest'' and an environment that is
ignored from an observational point of view, as required in EID, and
the resulting implication that the relevance of environmental
decoherence is restricted to local subsystems of the total (nonlocal)
quantum Universe, has been a subject of ongoing critical discussions
(see, for example, Refs.~\cite{Espagnat:1988:cf,Joos:2003:jh,%
  Zurek:1998:re,Schlosshauer:2003:tv}).  Furthermore, no general rule
is available that would indicate where the split between system and
environment is to be placed, a conceptual difficulty admitted also by
proponents of EID \cite[p.~1820]{Zurek:1998:re}.  These issues seem to
have motivated the attempt of the SID program to derive decoherence
for closed, undivided systems.

However, it is important to note that EID has clearly demonstrated
that the assumption of the existence of closed system is unrealistic
in essentially all cases \cite{Joos:1985:iu,Tegmark:1993:uz}.
Enlarging the system by including parts of its environment, as it is
implicitly done in SID in order to arrive at a quasicontinuous
spectrum, will render the closed-system assumption even less
physically viable: The combined system will in turn interact with its
surroundings, and the degree of environmental interaction will
increase with the number of degrees of freedom in the system. Also,
since some interaction with the external measuring device will be
required, the assumption of a closed system simply bypasses the
question of how the information contained in the ensemble is acquired
in the first place.  Ultimately, the only truly closed ``system'' is
the Universe in its entirety, and one can therefore question the
physical relevance and motivation for a derivation of decoherence for
subsystems that are presumed to be closed.

Furthermore, a general measurement in SID would pertain also to the
environment implicitly contained in the ``closed system,'' posing the
question of how this could translate into an experimentally realizable
situation. And even if such a measurement can be carried out, its
result would usually be of rather little physical interest in the
typical situation of observing decoherence for a particular object due
to its largely unobserved environment.

Finally, in SID, suppression of off-diagonal terms always occurs in
the energy eigenbasis, which can therefore be viewed as the universal
``preferred basis'' in this approach. However, this basis will
generally not be useful in accounting for our observation of different
preferred bases for the relevant local systems of interest (e.g.,
spatial localization of macroscopic bodies
\cite{Joos:2003:jh,Tegmark:1993:uz,Hornberger:2003:un,%
  Hornberger:2003:tv,Gallis:1990:un}, chirality eigenstates for
molecules such as sugar
\cite{Joos:1985:iu,Harris:1981:rc,Blanchard:2000:fq}, and energy
eigenstates in atoms \cite{Paz:1999:vv}). Furthermore, the energy
eigenbasis cannot be used to describe the emergence of time-dependent,
quasiclassical properties.

In conclusion, not only is the scope of SID more limited than that of
EID, but the two approaches also rest on different foundations. The
interpretation of the processes described by these theories is
fundamentally different, even though phenomenological effects of EID
can manifest themselves in a manner formally similiar to that of SID,
i.e., as a disappearance of off-diagonal terms in expectation values.
Any proposed derivations of an ``equivalence'' between SID and EID
\cite{Castagnino:2003:yb,Castagnino:2001:pl,Castagnino:2005:yb} can
therefore at most claim to describe coincidental formal similiarities
in the context of very particular models, and only if the scope of EID
is reduced to the influence on expectation values. On the basis of our
arguments, we question the justification for labeling the process
referred to by SID as ``decoherence.''

\section{\label{sec:zurek}Analysis of the spin bath model}

By studying an explicit model, we shall now directly investigate the
claim of SID, that terms not diagonal in energy in the expectation
value of arbitrary observables of the system decay if the system is
endowed with a continuous energy spectrum.  We shall also illustrate
formal and numerical differences in the time evolution of the
expectation value of local observables that take into account only the
degrees of freedom of the system $\mathcal{S}$ while ignoring the
environment $\mathcal{E}$ (the situation encountered in EID), and
global observables that pertain to both $\mathcal{S}$ and
$\mathcal{E}$ (the case treated by SID).  However, in view of our
arguments in the preceding Sec.~\ref{sec:differences}, this should not
be misunderstood as a side-by-side comparison of SID and EID. While
expectation values may share formal similiarities in both approaches,
they also obliterate fundamental differences between SID and EID that
lead to very different implications of these expectation values for
the question of decoherence.

\subsection{The model and its time evolution}

The probably most simple exactly solvable model for decoherence was
introduced some years ago by Zurek \cite{Zurek:1982:tv}. Here, the
system $\mathcal{S}$ consists of a spin-1/2 particle (a single qubit)
with two possible states $\ket{0}$ (representing spin up) and
$\ket{1}$ (corresponding to spin down), interacting with a collection
of $N$ environmental qubits (described by the states
$\ket{\uparrow_i}$ and $\ket{\downarrow_i}$) via the total Hamiltonian
\be \label{eq:hse-zurek}
\widehat{H}_\mathcal{SE} = \frac{1}{2} \bigl(\ketbra{0}{0} - \ketbra{1}{1}\bigr)
\sum_{i=0}^N g_i \bigl(\ketbra{\uparrow_i}{\uparrow_i} -
\ketbra{\downarrow_i}{\downarrow_i}\bigr)  \bigotimes_{i'\not= i} \,
\widehat{I}_{i'} .
\ee
Here, the $g_i$ are coupling constants, and $\widehat{I}_i =
(\ketbra{\uparrow_i}{\uparrow_i} +
\ketbra{\downarrow_i}{\downarrow_i})$ is the identity operator for the
$i$th environmental qubit.  The self-Hamiltonians of $\mathcal{S}$
and $\mathcal{E}$ are taken to be equal to zero. Note that
$\widehat{H}_\mathcal{SE}$ has a particularly simple form, since it
contains only terms diagonal in the $\{ \ket{0}, \ket{1} \}$ and $\{
\ket{\uparrow_i}, \ket{\downarrow_i} \}$ bases.

It follows that the eigenstates of $\widehat{H}_\mathcal{SE}$ are
product states of the form $\ket{\phi_\lambda} = \ket{0}
\ket{\uparrow_1} \ket{\downarrow_2} \cdots \ket{\uparrow_N}$, etc.  A
general state $\ket{\Psi_0}$ can then be written as a linear
combination of product eigenstates,
\be \label{eq;psi0-zurek}
\ket{\Psi_0} = \bigl(a\ket{0} + b\ket{1}\bigr) \bigotimes_{i=1}^N \, \bigl(\alpha_i
\ket{\uparrow_i} + \beta_i \ket{\downarrow_i}\bigr).
\ee
This state evolves under the action of $\widehat{H}_\mathcal{SE}$ into
\be \label{eq:psit-zurek}
\ket{\Psi(t)} = a\ket{0} \ket{\mathcal{E}_0(t)} + b\ket{1} \ket{\mathcal{E}_1(t)},
\ee
where
\be
\ket{\mathcal{E}_0(t)} = \ket{\mathcal{E}_1(-t)} = \bigotimes_{i=1}^N
\, \bigl(\alpha_i e^{ig_it/2}
\ket{\uparrow_i} + \beta_i e^{-ig_it/2} \ket{\downarrow_i}\bigr).
\ee
The density matrix is
\begin{multline}
  \hspace{-.2cm} \rho(t) = |a|^2
  \ket{0}\ket{\mathcal{E}_0(t)}\bra{\mathcal{E}_0(t)}\bra{0} + |b|^2
  \ket{1}\ket{\mathcal{E}_1(t)}\bra{\mathcal{E}_1(t)}\bra{1} \\ \, + ab^*
  \ket{0}\ket{\mathcal{E}_0(t)}\bra{\mathcal{E}_1(t)}\bra{1} + a^*b
  \ket{1}\ket{\mathcal{E}_1(t)}\bra{\mathcal{E}_0(t)}\bra{0},
\end{multline}
and its part diagonal in energy (\ie diagonal in the eigenstates
$\ket{\phi_\lambda}$ of $\widehat{H}_\mathcal{SE}$) is
\begin{widetext}
\be\label{eq:rho_d}
\begin{split}
\rho_d(t) &= \bigl(|a|^2\ketbra{0}{0} +
|b|^2 \ketbra{1}{1} \bigr) \underbrace{\bigl[
\hdots + |\alpha_1|^2 |\beta_2|^2
\cdots |\alpha_N|^2 \ket{\uparrow_1}\ket{\downarrow_2}\cdots \ket{\uparrow_N}
\bra{\uparrow_N} \cdots
\bra{\downarrow_2} \bra{\uparrow_1}
+ \hdots
\bigr]}_{\text{same-direction
  pairing}} \\
&+ \bigl(ab^* \ketbra{0}{1} +
a^*b \ketbra{1}{0} \bigr) \underbrace{\bigl[
\hdots + \beta_1 \alpha_1^* \beta_2
\alpha_2^* 
\cdots \alpha_N \beta_N^* \ket{\downarrow_1}\ket{\downarrow_2}\cdots \ket{\uparrow_N}
\bra{\downarrow_N} \cdots
\bra{\uparrow_2} \bra{\uparrow_1}
+ \hdots \bigr]}_{\text{opposite-direction
  pairing}}. 
\end{split}
\ee
\end{widetext}

\subsection{\label{sec:model-eid}Expectation values of local observables}

Focusing, in the spirit of EID, on the system $\mathcal{S}$ alone, we
trace out the degrees of freedom of the spin bath in the density
operator $\rho_\mathcal{SE} = \ketbra{\Psi(t)}{\Psi(t)}$.  This yields
the reduced density operator
\begin{multline} \label{eq:rho-s-zurek}
\rho_\mathcal{S} = \text{Tr}_\mathcal{E} \rho_\mathcal{SE} = |a|^2
\ketbra{0}{0} + |b|^2 \ketbra{1}{1} \\ + ab^* r(t) \ketbra{0}{1} + a^*b
r^*(t) \ketbra{1}{0},
\end{multline}
where the time dependence of the off-diagonal terms $\ketbra{0}{1}$
and $\ketbra{1}{0}$ is given by the decoherence factor 
\be  \label{eq:r}
r(t) = \braket{\mathcal{E}_1(t)}{\mathcal{E}_0(t)} = 
\prod_{i=1}^N \bigl(|\alpha_i|^2 e^{ig_it} + |\beta_i|^2 e^{-ig_it}\bigr).
\ee
The expectation value of any local $\mathcal{S}$ observable
\be \label{eq:o1}
\widehat{O}_\mathcal{S} = \biggl[ \sum_{s,s'=0,1} s_{ss'}
\ketbra{s}{s'} \biggr]
\bigotimes_{i=1}^N \, \widehat{I}_i,
\ee
is then given by
\bea \label{eq:o2}
\langle \widehat{O}_\mathcal{S} \rangle_{\Psi(t)} &=& \text{Tr}_\mathcal{SE} (\rho
\widehat{O}_\mathcal{S}) = \text{Tr}_\mathcal{S} (\rho_\mathcal{S}
\widehat{O}_\mathcal{S})\nonumber \\ &=& |a|^2 s_{00} + |b|^2 s_{11}
+ 2\, \text{Re} \bigl[ ab^* s_{10} r(t) \bigr]. 
\eea
We can formally rewrite $r(t)$ as a sum,
\be  \label{eq:r-int}
r(t) = \sum_\lambda |\braket{\Psi_0}{\phi_\lambda}|^2  e^{iE_\lambda t},
\ee
where the sum runs over all eigenstates $\ket{\phi_\lambda}$ of the total
Hamiltonian $\widehat{H}_\mathcal{SE}$, with eigenvalues $E_\lambda$.

\begin{figure}
\begin{center}
\includegraphics[scale=0.63]{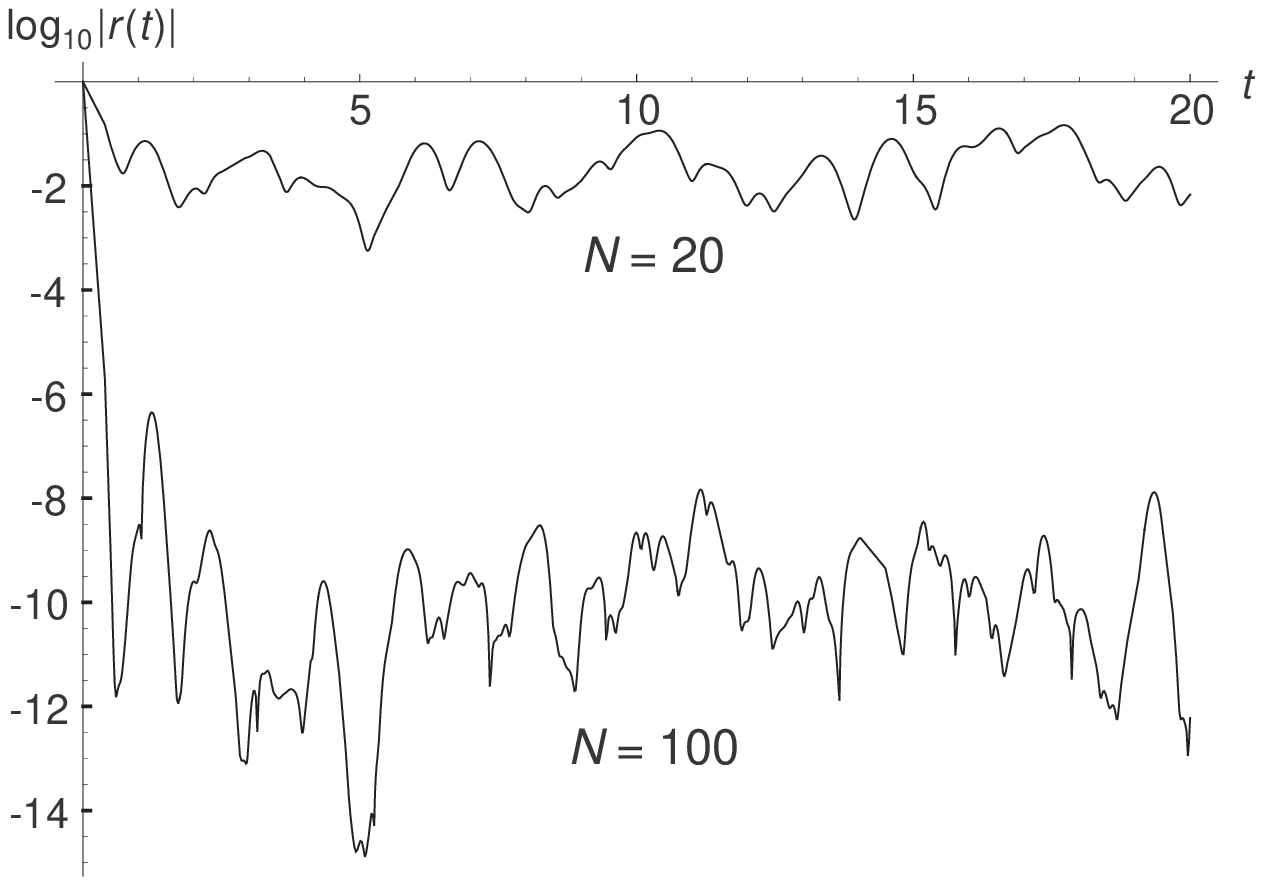}
\caption[Plot of  the
decoherence factor $r(t)$]{\label{fig:r-eid}Plot of $\log_{10}
  |r(t)|$, with the decoherence factor $r(t)$ given by
  Eq.~\eqref{eq:r}, for two different bath sizes $N=20$ and $100$.
  Fast decay of $r(t)$, corresponding to local decoherence, is
  observed, and the degree of decoherence is seen to increase with
  $N$. The squared coefficients $|\alpha_i|^2$ and the couplings $g_i$
  were drawn from a uniform random distribution over the intervals
  $[0,1]$ and $[-\pi,\pi]$, respectively.}
\end{center}
\end{figure}

A concrete illustration for the time dependence of $r(t)$,
Eq.~\eqref{eq:r}, for two different bath sizes is shown in
Fig.~\ref{fig:r-eid}. We see that $|r(t)|$ decays quickly by several
orders of magnitude and then continues to oscillate about a very small
mean value. Thus, for local observables, terms corresponding to
interference between the two $\mathcal{S}$ states $\ket{0}$ and
$\ket{1}$ become quickly and strongly suppressed.

\subsection{\label{sec:model-sid}Expectation values of global observables}

An arbitrary global observable $\widehat{O}\equiv
\widehat{O}_\mathcal{SE}$ can be written as a linear combination of
the form $\widehat{O} = \sum_{\lambda \lambda'} O_{\lambda \lambda'}
\ket{\phi_\lambda} \bra{\phi_{\lambda'}}$, where the
$\ket{\phi_\lambda}$ are product eigenstates of the total Hamiltonian
$\widehat{H}_\mathcal{SE}$, Eq.~\eqref{eq:hse-zurek}. Explicitly,
\begin{multline} \label{eq:o-general}
\widehat{O} =  \sum_j \bigl( s^{(j)}_{00} \ketbra{0}{0} +
s^{(j)}_{01} \ketbra{0}{1} + s^{(j)}_{10}
\ketbra{1}{0} + s^{(j)}_{11} \ketbra{1}{1} \bigr)
 \\ \,\,\,  \bigotimes_{i=1}^N \,  \bigl( \epsilon^{(ij)}_{\uparrow\uparrow}
\ketbra{\uparrow_i}{\uparrow_i} + \epsilon^{(ij)}_{\uparrow\downarrow}
\ketbra{\uparrow_i}{\downarrow_i} + \epsilon^{(ij)}_{\downarrow\uparrow}
\ketbra{\downarrow_i}{\uparrow_i} + \epsilon^{(ij)}_{\downarrow\downarrow}
\ketbra{\downarrow_i}{\downarrow_i} \bigr).
\end{multline}
Since $\widehat{O}$ must be Hermitian, $s_{00}$, $s_{11}$,
$\epsilon^{(i)}_{\uparrow\uparrow}$, and
$\epsilon^{(i)}_{\downarrow\downarrow}$ are real numbers, and $s_{01}
= \bigl( s_{10} \bigr)^*$, $\epsilon^{(i)}_{\downarrow\uparrow} =
\bigl( \epsilon^{(i)}_{\uparrow\downarrow} \bigr)^*$.  To keep the
notation simple, we shall omit the sum over $j$ (and thus the index
$j$) in the following.

The expectation value of $\widehat{O}$ in the state $\ket{\Psi(t)}$,
Eq.~\eqref{eq:psit-zurek}, is 
\begin{widetext}
\bea \label{eq:exptval} 
\langle \widehat{O} \rangle_{\Psi(t)} &=&
  \bigl( |a|^2 s_{00} + |b|^2 s_{11} \bigr) \prod_{i=1}^N \bigl[ |\alpha_i|^2
  \epsilon^{(i)}_{\uparrow\uparrow} + \alpha_i^*
  \beta_i \epsilon^{(i)}_{\uparrow\downarrow} e^{-ig_it} + \bigl(\alpha_i^*
  \beta_i \epsilon^{(i)}_{\uparrow\downarrow}\bigr)^* e^{ig_it} +
  |\beta_i|^2
  \epsilon^{(i)}_{\downarrow\downarrow} \bigr] \nonumber \\
  && + \, 2 \, \text{Re} \biggl( ab^* s_{10} \prod_{i=1}^N \bigl[
  |\alpha_i|^2 \epsilon^{(i)}_{\uparrow\uparrow} e^{ig_it}+ 
  \alpha^*_i \beta_i
  \epsilon^{(i)}_{\uparrow\downarrow}  + 
  \bigl( \alpha_i^* \beta_i
   \epsilon^{(i)}_{\uparrow\downarrow}\bigr)^*  + |\beta_i|^2
  \epsilon^{(i)}_{\downarrow\downarrow} e^{-ig_it} \bigr] \biggr) \nonumber \\
  &\equiv & \bigl( |a|^2 s_{00}  + |b|^2 s_{11} \bigr) \Gamma_{0}(t) +
  2 \, \text{Re} \bigl[ ab^* s_{10} \Gamma_{1}(t) \bigr].  
\eea
\end{widetext}
The special case of the expectation value of local observables, as
considered in the preceding Sec.~\ref{sec:model-eid}, can easily be
recovered by remembering that tracing out the degrees of freedom of
$\mathcal{E}$ is equivalent to choosing all coefficients
$\epsilon^{(i)}_{\uparrow\uparrow}=\epsilon^{(i)}_{\downarrow\downarrow}=1$
and $\epsilon^{(i)}_{\uparrow\downarrow} =
\bigl(\epsilon^{(i)}_{\downarrow\uparrow}\bigr)^* = 0$, which yields
$\Gamma_{0}(t) = 1$ and $\Gamma_{1}(t) = r(t)$ [see Eq.~\eqref{eq:r}],
in agreement with Eq.~\eqref{eq:o2}.

Suppression of terms in $\langle \widehat{O} \rangle_{\Psi(t)}$ that
are not diagonal in the energy eigenbasis would be represented by the
vanishing of all time-dependent terms in the above expression, \ie
\bea \label{eq:o_d} 
\langle \widehat{O} \rangle_d &=& \bigl( |a|^2 s_{00}
+ |b|^2 s_{11} \bigr) \prod_{i=1}^N \big( |\alpha_i|^2 \epsilon^{(i)}_{\uparrow\uparrow} +
|\beta_i|^2
\epsilon^{(i)}_{\downarrow\downarrow} \big) 
\nonumber \\ && \, + 2 \, \text{Re} \biggl( ab^* s_{10} \prod_{i=1}^N 2 \,
\text{Re} \bigl( \alpha^*_i \beta_i
\epsilon^{(i)}_{\uparrow\downarrow}   \bigr) \biggr) \nonumber \\
&\equiv & \bigl( |a|^2 s_{00}  + |b|^2 s_{11} \bigr) \Gamma^d_{0}  + 
2 \, \text{Re} \bigl( ab^* s_{10} \Gamma^d_{1} \bigr), 
\eea
because we can easily show that $\langle \widehat{O} \rangle_d =
\text{Tr} (\rho_d \widehat{O} )$, where $\rho_d$,
Eq.~\eqref{eq:rho_d}, is the part of the density matrix that is
diagonal in the eigenstates of the total Hamiltonian. We also see that
$\langle \widehat{O} \rangle_d = \text{Tr} (\rho \widehat{O}_d )$,
where
\be \label{eq:od}
\begin{split}
\widehat{O}_d &= \bigl( s_{00} \ketbra{0}{0} +
s_{11} \ketbra{1}{1} \bigr) \bigotimes_{i=1}^N \, \bigl( \epsilon^{(i)}_{\uparrow\uparrow}
\ketbra{\uparrow_i}{\uparrow_i} + \epsilon^{(i)}_{\downarrow\downarrow}
\ketbra{\downarrow_i}{\downarrow_i} \bigr) 
\\ & + \bigl( s_{01} \ketbra{0}{1} + s_{10}
\ketbra{1}{0} \bigr) 
\bigotimes_{i=1}^N \, \bigl( \epsilon^{(i)}_{\uparrow\downarrow}
\ketbra{\uparrow_i}{\downarrow_i} + \epsilon^{(i)}_{\downarrow\uparrow}
\ketbra{\downarrow_i}{\uparrow_i} \bigr).
\end{split}
\ee
is the part of $\widehat{O}$ diagonal in energy. Thus, as expected,
diagonality of $\langle \widehat{O} \rangle_{\Psi(t)}$ in energy can
also be characterized by the presence of only those product expansion
coefficients that are contained in $\widehat{O}_d$.

The form of the two product terms $\Gamma_{0}(t)$ and $\Gamma_{1}(t)$
is similar: They only differ in the order of the pairing of the
product expansion coefficients with the exponential factors. Also,
since the coefficients $s_{jj'}$ are independent, diagonalization in
energy will in general require that individually $\Gamma_{j}(t)
\rightarrow \Gamma_{0}^d$ and $\Gamma_{1}(t) \rightarrow \Gamma_{1}^d$
for large $t$. We can therefore restrict our following analysis to
$\Gamma_{0}(t)$ alone. (We shall also omit the subscript ``0'' in the
following.)

First of all, let us rewrite $\Gamma(t)$ as a sum of $4^N$ terms,
\be \label{eq:gamma-1}
\Gamma(t) = \sum_{\lambda} c_\lambda e^{iE_\lambda t}.
\ee
where the $c_\lambda$ represent products of expansion coefficients,
\bea 
\label{eq:c-lambda}
c_\lambda &=& \biggl( \prod_{i \in I_1(\lambda)} |\alpha_i|^2
\epsilon^{(i)}_{\uparrow\uparrow} \biggr) \biggl( \prod_{i \in
  I_2(\lambda)} |\beta_i|^2 \epsilon^{(i)}_{\downarrow\downarrow}
\biggr) \nonumber \\ &\times & \biggl( \prod_{i \in I_3(\lambda)}
\alpha_i^* \beta_i \epsilon^{(i)}_{\uparrow\downarrow} \biggr) \biggl(
\prod_{i \in I_4(\lambda)} \bigl( \alpha^*_i \beta_i
\epsilon^{(i)}_{\uparrow\downarrow}\bigr)^* \biggr).
\eea
Here the sets $I_k(\lambda)$ specify over which indices $i$ each
product runs, namely, they are subsets of the set $I=\{1,\hdots,N\}$
of all integers between 1 and $N$ such that $\cup_k I_k(\lambda) = I$
and $\cap_k I_k(\lambda) = \emptyset $. The total energy $E_\lambda$
associated with each term in the sum, Eq.~\eqref{eq:gamma-1}, is
\be
E_\lambda = \sum_{i \in I_4(\lambda)} g_i - \sum_{i \in I_3(\lambda)} g_i. 
\ee
We choose the index $\lambda$ such that $E_{\lambda-1} \le E_\lambda
\le E_{\lambda+1}$ for all $\lambda$. Clearly, $E_\lambda = 0$
whenever $I_3(\lambda) = I_4(\lambda) = \emptyset$ [\ie if
$I_1(\lambda) \cup I_2(\lambda) = I$], canceling out the time
dependence of the associated product term in the expression for
$\Gamma(t)$.  Thus, we can split $\Gamma(t)$ into a
time-independent and a time-dependent part,
\bea \label{eq:gamma-int}
\Gamma(t) &=& \sum_{\lambda} c_\lambda
+ \sum_\lambda c_\lambda e^{iE_\lambda t}
\equiv \Gamma^d + \Lambda(t),
\eea
where now the first sum runs over all $\lambda$ for which
$I_1(\lambda) \cup I_2(\lambda) = I$, while the second sum runs over
all $\lambda$ for which $ I_3(\lambda) \cup I_4(\lambda) \not=
\emptyset$.

Diagonality in energy would require $\Lambda(t) \rightarrow 0$ as
$t \rightarrow \infty$. Written this way, we see that $\Lambda(t)$ is
formally similiar to the function $r(t)$ derived for local
observables, Eq.~\eqref{eq:r-int}.  This might not come as a surprise,
since also the expression for $r(t)$ can be derived from the
calculation of an expectation value of an observable, namely, that of
the local observable $\widehat{O}_r = (\ketbra{0}{1} + \ketbra{1}{0})
\bigotimes_{i=1}^N \, \widehat{I}_k$ that measures the degree of local
interference between the $\mathcal{S}$ states $\ket{0}$ and $\ket{1}$.
However, in the case of $r(t)$, $c_\lambda = |
\braket{\phi_\lambda}{\Psi_0} |^2$ is a product of $N$ real and
non-negative coefficients $|\alpha_i|^2$ and
$|\beta_i|^2=1-|\alpha_i|^2$, while the $c_\lambda$ of
Eq.~\eqref{eq:c-lambda} contain cross terms of the form $\alpha_i
\beta_i^*$ and $\alpha_i^* \beta_i$, arbitrary real coefficients
$\epsilon^{(i)}_{\uparrow\uparrow}$ and
$\epsilon^{(i)}_{\downarrow\downarrow}$, and arbitrary complex
coefficients $\epsilon^{(i)}_{\uparrow\downarrow}$.

We expect this difference to have strong influence on the time
evolution of $\Lambda(t)$ vs that of $r(t)$. The destructive
interference needed to obtain suppression of the off-diagonal part of
the expectation value relies on the idea that, when a function $f(z)$
is multiplied by a phase factor $e^{izt}$ whose variation with $z$ is
much faster than that of $f(z)$, neighboring values $f(z)$ and
$f(z+\delta z)$ will have similiar magnitude and phases, but will be
weighted with two strongly different phase factors, which leads to an
averaging-out effect in the sum $\sum_z f(z)e^{izt}$.

In our case, writing 
\be\label{eq:gamma-int2}
\Lambda(t) = \sum_\lambda  r_\lambda e^{i\varphi_\lambda} e^{i E_\lambda t},
\ee
with $r_\lambda = |c_\lambda|$, the phases $\varphi_\lambda$ will in
general vary very rapidly with $\lambda$ and, thus, with $E_\lambda$.
This is a consequence of the fact that the $c_\lambda = r_\lambda
e^{i\varphi_\lambda} $ are composed of products of coefficients, such
that changing a single term in the product will in general result in a
drastic change in the overall phase associated with the $c_\lambda$.
(The variation in magnitude among the $c_\lambda$ can be expected to
be comparably insignificant for larger $N$.) Such discontinuous phase
fluctuations are absent in the formally similiar function $r(t)$,
Eq.~\eqref{eq:r}, since there only the absolute value of the
coefficients $\alpha_i$ and $\beta_i$ enters.  Note that the impact of
the phase fluctuations cannot be diminished by going to larger $t$,
since the $2\pi$ periodicity of phases implies that the effect of a
phase difference between terms $\lambda$ and $\lambda + 1$ induced by
$e^{i(E_{\lambda+1} - E_\lambda) t}$ will in average be similiar to
that induced by $e^{i (\varphi_{\lambda+1} - \varphi_\lambda)}$ for
all (larger) values of $t$.

We anticipate the described phase-variation effect to counteract the
averaging-out influence of the multiplying phase factor $e^{-iEt}$, and
to thus make it more difficult, if not entirely impossible, for
$\Lambda(t)$, Eq.~\eqref{eq:gamma-int}, to converge to zero.  On the
other hand, if the average difference between the phases associated
with the individual coefficients is decreased, we would expect that
the rate and degree of decay of $\Lambda(t)$ will be improved.

\subsection{Numerical results for the expectation value of random global observables}

To check this prediction and to generally gain more insight into the
behavior of $\langle \widehat{O} \rangle_{\Psi(t)}$,
Eq.~\eqref{eq:exptval}, we studied numerically the time evolution of
$\Lambda(t)$, Eq.~\eqref{eq:gamma-int2}, normalized by its initial
value at $t=0$, for sets of random observables $\widehat{O}$.
Diagonalization of $\langle \widehat{O} \rangle_{\Psi(t)}$ in energy
would then be represented by a decay of $\Lambda(t)$ from its initial
value of one.

\begin{figure}
\begin{center}
\includegraphics[scale=.8]{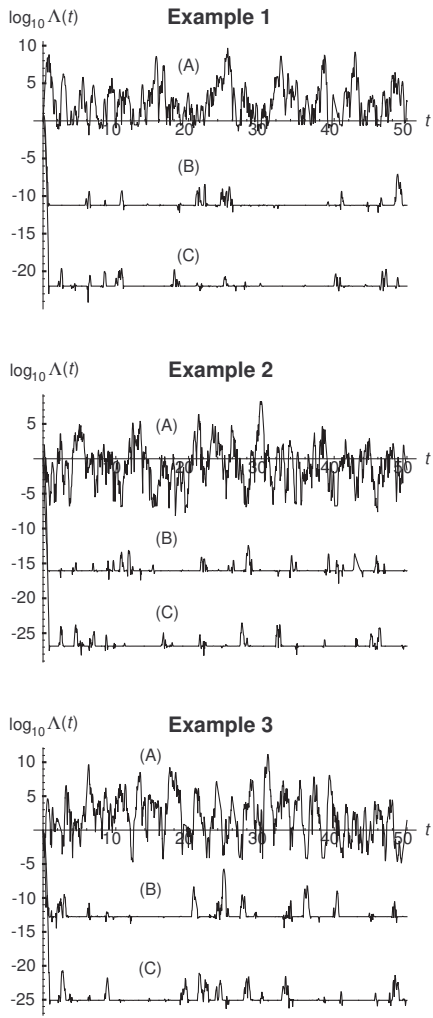}
\caption[Time evolution of $\log_{10} \Lambda(t)$ for $N=100$ bath spins and  three
different random observables and initial states of the
environment]{\label{fig:q}Time evolution of $\log_{10} \Lambda(t)$
  [see Eq.~\eqref{eq:gamma-int2}] for $N=100$ bath spins and three
  different random observables and initial states of the environment.
  The function $\Lambda(t)$ quantifies the time dependence of the
  terms in the expectation value $\langle \widehat{O}
  \rangle_{\Psi(t)}$ [Eq.~\eqref{eq:exptval}] that are not diagonal in
  energy. Suppression of these terms is represented by a decay of
  $\Lambda(t)$ from its initial value of one [i.e., $\log_{10}
  \Lambda(t) \longrightarrow -\infty$].  It is observed that in the
  general case (A) of completely random observables and initial states
  of the environment, collective decay of off-diagonal terms does, in
  general, not occur.  However, if the phases of the coefficients
  describing the observable and the environment are moderately
  restricted (B) or all set equal to zero (C), decay of increasing
  strength is found. }
\end{center}
\end{figure}

Figure~\ref{fig:q} shows three typical examples for the time evolution
of $\log_{10} \Lambda(t)$ for a fixed bath size of $N=100$. All couplings $g_i$
were taken to be random real numbers between $-\pi$ and $\pi$. To
investigate the influence of phase fluctuations of the $c_\lambda$,
Eq.~\eqref{eq:c-lambda}, we considered three different cases for
selecting the coefficients $\alpha_i$, $\beta_i$,
$\epsilon^{(i)}_{\uparrow\uparrow}$,
$\epsilon^{(i)}_{\downarrow\downarrow}$, and
$\epsilon^{(i)}_{\uparrow\downarrow}$, i.e., for choosing the initial
state of the environment and the observable. In the completely random
case (A), the coefficients $\alpha_i$, $\beta_i$, and
$\epsilon^{(i)}_{\uparrow\downarrow}$ were taken to be random complex
numbers, with magnitudes and phases drawn from a uniform distribution
over the intervals $[0,1]$ and $[0,2\pi]$, respectively (and such that
$|\beta_i|^2 = 1 - |\alpha_i|^2$). Similiarly, the coefficients
$\epsilon^{(i)}_{\uparrow\uparrow}$ and
$\epsilon^{(i)}_{\downarrow\downarrow}$ were random real numbers drawn
from a uniform distribution over the interval $[-1,1]$. In the second
case (B), the initial state of the environment was prepared such that
the phases of the $\alpha_i$ and $\beta_i$ were restricted to the
interval $[0, \pi/2]$. Also, only observables with non-negative values
of $\epsilon^{(i)}_{\uparrow\uparrow}$ and
$\epsilon^{(i)}_{\downarrow\downarrow}$ were considered, such that
sign reversals of $c_\lambda$ due to a change of product terms
containing these coefficients were prevented. Finally, in the third
case (C), only the absolute values of the $\alpha_i$, $\beta_i$,
$\epsilon^{(i)}_{\uparrow\uparrow}$,
$\epsilon^{(i)}_{\downarrow\downarrow}$, and
$\epsilon^{(i)}_{\uparrow\downarrow}$ were used, which implies that
the $c_\lambda$ fluctuated only in magnitude.
    
We observed a drastic influence of the range of phases and signs
associated with the individual coefficients $\alpha_i$, $\beta_i$,
$\epsilon^{(i)}_{\uparrow\downarrow}$,
$\epsilon^{(i)}_{\uparrow\uparrow}$, and
$\epsilon^{(i)}_{\downarrow\downarrow}$, on the evolution of
$\Lambda(t)$.  In the special case (C) of all coefficients being real
non-negative numbers, $\Lambda(t)$ exhibited a consistently strong and
fast decay behavior, similiar to the decay of the function $r(t)$,
Eq.~\eqref{eq:r}, describing suppression of off-diagonal terms for
local observables (see Fig.~\ref{fig:r-eid}). In the intermediate case
(B), with restricted phases and signs, the degree of decay of
$\Lambda(t)$ was decreased, while the decay rate stayed roughly the
same.  In the general random case (A), in which no restriction on the
spread of phases and on the signs of the coefficients was imposed, the
time evolution of $\Lambda(t)$ was observed to be sensitive to the
particular set of random numbers used for the coefficients $\alpha_i$,
$\beta_i$, $\epsilon^{(i)}_{\uparrow\downarrow}$,
$\epsilon^{(i)}_{\uparrow\uparrow}$, and
$\epsilon^{(i)}_{\downarrow\downarrow}$ in each run.  For some of the
sets, $\Lambda(t)$ was seen to lack any decay behavior at all. In
other cases, the baseline of oscillation was located below zero,
indicating a very weak damping effect, albeit with the peaks of
the large-amplitude oscillation frequently reaching values greater
than zero.

These results show that, for the bath size studied here, a consistent
occurence of a decay of $\Lambda(t)$ hinges on the phase restrictions
imposed on the coefficients describing the observable and the initial
state of the environment.  If these restrictions are given up, the
time evolution of $\Lambda(t)$ and any occurence of a (comparably
weak) decay will exhibit strong dependence on the particular set of
values chosen for the coefficients.

\begin{figure}
\begin{center}
\includegraphics[scale=0.6]{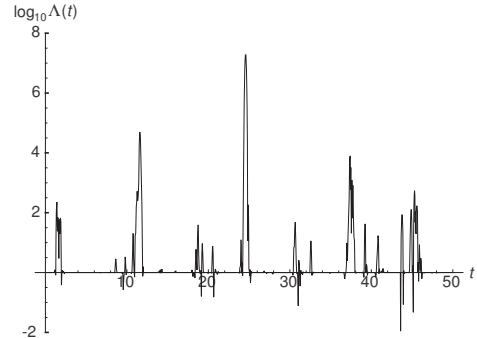} \\[.5cm]
\caption[Time evolution of $\log_{10} \Lambda(t)$ for
$N=100$ bath spins when no restrictions on the initial state of the
environment are imposed]{\label{fig:q2}Time evolution of $\log_{10}
  \Lambda(t)$ for $N=100$ bath spins [see Eq.~\eqref{eq:gamma-int2}]
  when no restrictions on the initial state of the environment are
  imposed.  Such restrictions are physically unrealistic and require a
  preparing measurement on the unrestricted environment, which would
  in turn be in conflict with the desired generality of the derivation
  of decay effects. No collective decay of off-diagonal terms is
  observed, regardless of any restrictions imposed on the observable.
}
\end{center}
\end{figure}

However, it is important to realize that the assumption of a
restricted initial state of $\mathcal{E}$ is not only unrealistic,
since the environment is typically uncontrollable, but it will also
lead to a circular argument when aiming at a derivation of a universal
decay effect. This is so because any restriction would require an
appropriate preparation of the initial state through a measurement on
the entire $\mathcal{E}$, which implies that suppression of
off-diagonal terms would then in general be absent for the observable
corresponding to this measurement, if the restriction of the initial
state of $\mathcal{E}$ is relevant to the occurence of the
suppression.  Consequently, the $\alpha_i$ and $\beta_i$ must be
allowed to possess arbitrary phases.  Then, since the
$\epsilon^{(i)}_{\uparrow\uparrow}$,
$\epsilon^{(i)}_{\downarrow\downarrow}$, and
$\epsilon^{(i)}_{\uparrow\downarrow}$ are always paired with the
$\alpha_i$ and $\beta_i$ in the expression for the $c_\lambda$ that
make up $\Lambda(t)$ [see Eq.~\eqref{eq:c-lambda}], we anticipate that
giving up phase restrictions on the $\alpha_i$ and $\beta_i$ will
render the restrictions imposed on the $\widehat{O}$-coefficients less
effective, if not entirely irrelevant, in bringing about a decay of
$\Lambda(t)$.

To study this prediction, in Fig.~\ref{fig:q2} we show a
representative plot of $\Lambda(t)$ using only the absolute values of
the $\widehat{O}$ coefficients $\epsilon^{(i)}_{\uparrow\uparrow}$,
$\epsilon^{(i)}_{\downarrow\downarrow}$, and
$\epsilon^{(i)}_{\uparrow\downarrow}$, but with the
$\mathcal{E}$ coefficients $\alpha_i$ and $\beta_i$ possessing random
phases between 0 and $2\pi$. We found that decay is either entirely
absent or strongly diminished in strength, despite the fact that the
strongest possible restriction on the phases and signs of the
$\widehat{O}$ coefficients is imposed. Similiar to the case of
completely random coefficients, the behavior of $\Lambda(t)$ was
observed to depend crucially on the particular set of random numbers
chosen for the coefficients. These results lead us to conclude that a
universal decay of off-diagonal terms does not occur for the studied
bath size and time scale.

\begin{figure}
\begin{center}
\includegraphics[scale=.69]{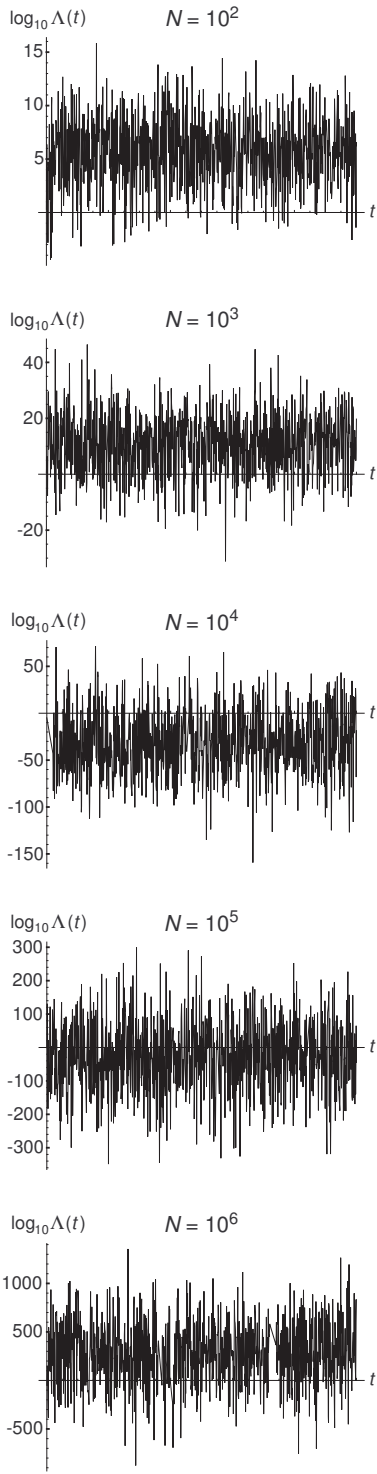}
\caption[Example for the time evolution of $\log_{10} \Lambda(t)$
using a random observable and random initial bath state, for bath
sizes between $N=10^2$ and $10^6$ and a long timescale
$t=0..10^6$]{\label{fig:q-lim}Example for the time evolution of
  $\log_{10} \Lambda(t)$ [see Eq.~\eqref{eq:gamma-int2}] using a
  random observable and random initial bath state, for bath sizes
  between $N=10^2$ and $10^6$ and a long time scale $t=0$--$10^6$.  No
  connection between the size of the spin bath and the occurrence and
  the degree of damping is observed.  Therefore, no consistent
  collective decay of interference terms occurs. The increased time
  scale is seen to be irrelevant.}
\end{center}
\end{figure}

To be sure, SID is based on the assumption of a quasicontinuous energy
spectrum and very long time scales, corresponding to ``sufficiently
large'' $N$ and $t$
(the existing derivations of SID \cite{Castagnino:2004:tc,Castagnino:2003:yb,Castagnino:2002:ic,%
Castagnino:2001:pq,Castagnino:2001:pl,Castagnino:2000:oq,%
Castagnino:2000:pb,Castagnino:2000:ja,Laura:1999:on,%
Castagnino:2000:vt,Castagnino:1998:tz,Laura:1998:io,Castagnino:1997:wz}
even assume the strict limits $N \rightarrow \infty$ and $t
\rightarrow \infty$, in order to allow for a direct application of the
Riemann-Lebesgue theorem), while so far we have only considered
relatively modest values for these parameters. However, since we know
from Fig.~\ref{fig:r-eid} that for expectation values of local
observables, strong and fast decay of off-diagonal terms is obtained
for the value of $N$ and over the time scale used in the plots shown in
Fig.~\ref{fig:q}, it is clear that, if a general global disappearance
of interference terms is to occur in our model, it will require a much
larger number of environmental qubits and/or longer time scales than
typically considered for local observables.

Accordingly, in Fig.~\ref{fig:q-lim} we show a typical example for the
time evolution of $\Lambda(t)$ over the time scale $t=0$--$10^6$ for the
case of a completely random observable and initial state of
$\mathcal{E}$, using comparably large bath sizes $N$ between $N=10^2$
and $10^6$.  We observed that even for these values of $N$, no
consistent occurrence of a decay became apparent.  In particular, no
generally valid direct correlation between the value of $N$ and the
time evolution of $\Lambda(t)$ was visible.  Instead, it was again the
particular set of random numbers included in the computation of
$\Lambda(t)$ for a given value of $N$ (but not to the size $N$ of the
set itself) that determined whether the baseline of oscillation of
$\Lambda(t)$ was located above or below the zero line.  In agreement
with analytical predictions in the preceding section, we also found
that the choice of a longer timescale is irrelevant, since neither the
baseline nor the amplitude of oscillation changed significantly over
the investigated time interval after a comparably short initial
period. Furthermore, we observed that even if $\Lambda(t)$ ``decayed''
for a particular set of random numbers, the function sustained a
large-amplitude oscillation whose peaks often attained values much
larger than the initial value of $\Lambda(t)$.

Our results show that, in general, for the bath sizes and time scales
studied, destructive interference of off-diagonal terms in the
expectation value expressed in the energy eigenbasis [as quantified by
$\Lambda(t)$, see Eq.~\eqref{eq:gamma-int2}] does not occur in our
model.  Instead, the time evolution of $\Lambda(t)$ is simply
determined by the particular random numbers used to describe the
observable and the initial state of the environment. Therefore, no
general suppression of interference terms can be inferred.

\section{\label{sec:discussion}Discussion}

The process described by SID appears to be neither formally nor
conceptually nor physically related to the decoherence mechanism in
the standard sense of environmental decoherence. EID accounts for the
absence of interference from the perspective of the local (open)
system by describing interactions with an environment in
quantum-mechanical terms of wave-function entanglement. In contrast,
SID describes dynamically induced destructive interference between
time-dependent terms in the expression for expectation values. SID
does not, however, explain the physical origin of the measurement
outcomes and their probability-weighted ensembles needed to define the
expectation values. Even if this purely phenomenological basis of SID
is accepted, the described process has no bearing on a loss of
coherence in individual measurements, since it is only a consequence
of averaging over a large number of measurement results. This is in
fundamental contrast to EID, where each measurementlike interaction
leads to a dislocalization of interference and thus, locally, to a
disappearance of interference.

The main result of our study of the spin bath model is the finding
that the destructive interference predicted by SID will in general
fail to occur in our model even for bath sizes and over time scales
much larger than typically considered in treatments of the same model
in environmental decoherence.  The source of this failure lies in the
random relative phases associated with the individual initial bath
spin states and the expansion coefficients of the observable. The
resulting discontinuous phase fluctuations in the coefficient function
$c_\lambda$, as defined in Eq.~\eqref{eq:c-lambda}, counteract the
supposed averaging-out effect of the dynamical phase factors $e^{iEt}$
in a way that is, due to the $2\pi$ periodicity of the phase,
effectively independent of the value of $t$.

Even when the bath size is increased, the function $c_\lambda$ remains
a set of discrete values with discontinuously varying phases. This can
be explained by noting that, while the total energy is a sum of the
energies of each subsystem, such that enlarging the number of
contributing subsystems will in general lead to an improved
quasicontinuity of the energy spectrum, the $2\pi$ periodicity of the
phases implies that the degree of phase discontinuity of the
$c_\lambda$ will not be diminished by increasing the number of
subsystems. It is therefore unlikely that a consistent decay behavior
could become apparent for spin baths much larger than those considered
here.

This indicates that it is not the degree of continuity of the energy
spectrum that represents the determining factor for obtaining
destructive interference.  Rather, it is the discrete nature of the
model itself that seems to lead to difficulties.  Only if restrictions
are imposed on both the measured observable and the initial state of
the environment, a consistent and general suppression of off-diagonal
terms can occur.  But, as we have argued, the corresponding
preparation of the initial state of the environment is physically
unrealistic and renders the derivation of a universal decay effect
circular. 

We conjecture that the diagonalization of the expectation value, as
described by SID, is likely to fail also in other systems composed of
discrete individual subentities.  For, in such models, the relevant
function will typically be represented by a large product of discrete
expansion coefficients, similiar to the $c_\lambda$ of our model,
whose discontinuous phase fluctuations will again be likely to
counteract the averaging-out influence of the dynamical phases.  It is
therefore clear that the seemingly innocuous mathematical requirement
of regularity and integrability of the coefficient functions (see
Sec.~\ref{sec:basic-concepts}) is far from ``valid in all relevant
cases'' where the condition of a sufficiently continuous energy
spectrum holds. The suggestion to approximate such discrete functions
by a continuous function through interpolation
\cite{Castagnino:2004:tc} does not appear to be viable, since the
interpolated function would describe a physically different situation.

On a general note, it is also important to realize that dynamical
phases are correlated. Thus one could always construct an observable
for which the initial phases of the coefficients seem completely
random, but are in fact chosen such that recurrence of coherence will
show up within a finite time interval, thus disproving the claimed
universality of SID without any further argument.

\section{\label{sec:conclusion}Summary and conclusions}

We have investigated the two main claims of the ``self-induced
decoherence'' approach, namely, (1) that expectation values of
observables pertaining to a closed system become diagonal in the
eigenbasis of the Hamiltonian, provided the system is endowed with a
continuous energy spectrum; and (2) that this process represents a
new way of describing quantum decoherence, and that it leads to
results equivalent to the standard approach of environment-induced
decoherence.

We have evaluated the first claim in the context of a simple spin bath
model of finite size by studying, analytically and numerically, the
time evolution of expectation values of random global observables. We
have found that, in general, collective decay of terms off-diagonal in
the energy eigenbasis does not occur over the large range of bath
sizes and time scales considered.  This result is not due to an
insufficient quasicontinuity of the energy spectrum, but is rather
rooted in the randomness of the phases associated with the observable
and the initial state of the environment. Even in the limit of large
bath sizes, the discrete functions for which destructive interference
is to be derived do not approach their sufficiently smoothly varying
interpolated approximations required for the dynamical phase averaging
to have an effect.

These results represent an example for a simple model system that,
although endowed with a quasicontinuous energy spectrum, fails to
exhibit the decay of off-diagonal terms that would be expected from an
extrapolation of SID to discrete models in the limit of comparably
large sizes of the system.  Such an extrapolation should be possible
if the approach is to have general physical relevance.  We have also
anticipated that the decay effect described by SID will likely be
absent also in other similiar models that are composed of discrete
subsystems.

With respect to the second claim of the SID program, we have
questioned the suggestion that SID represents a ``new viewpoint''
\cite{Castagnino:2004:tc} on the theory of environment-induced
decoherence, since the two approaches are based on conceptually,
formally, and physically unrelated mechanisms. In particular, we have
pointed out the following key differences and objections.

(i) SID does not describe the suppression of interference for
individual measurements, since interference terms in the expectation
value are not damped individually.
  
(ii) SID simply presupposes the existence of an ensemble of
measurement outcomes, without giving an account of its origin in terms
of a physical description of measurement.
  
(iii) The assumption of closed systems is unrealistic, especially for
systems containing the many degrees of freedom needed to obtain the
required quasicontinuous energy spectrum.
  
(iv) The physical feasibility and relevance of measurements pertaining
to the total system-environment combination is doubtful.
  
(v) Energy as the universal preferred basis of the global closed
system can usually not account for the different observed preferred
bases for the local system of interest.

Our study leads us to two main conclusions. First, it points to the
need for more precise, physically motivated criteria for the occurrence
of the destructive interference effect described by SID.  Most
importantly, however, the physical interpretation and relevance of
this effect need to be explained. We suspect that the SID approach may
have mistakenly interpreted and labeled an unrelated process as
``decoherence.''

\begin{acknowledgments}
  
The author would like to thank A.~Fine, M.~Castagnino, E.~Joos,
O.~Lombardi, and H.\ D.~Zeh for many thoughtful comments and helpful
discussions.

\end{acknowledgments}

\bibliographystyle{apsrev}

\bibliography{references}

\end{document}